\documentclass[twocolumn,showpacs,preprintnumbers,amsmath,amssymb, showkeys]{revtex4}

\usepackage{graphicx}
\usepackage{dcolumn}
\usepackage{bm}
\usepackage{epsf}

\begin{document}


\title{Tunneling transport through multi-electrons states in coupled quantum dots with Coulomb correlations}

\author{V.\,N.\,Mantsevich}
 \altaffiliation{vmantsev@spmlab.phys.msu.ru}
\author{N.\,S.\,Maslova}%
 \email{spm@spmlab.phys.msu.ru}
\author{P.\,I.\,Arseyev}
 \altaffiliation{ars@lpi.ru}
\affiliation{Moscow State University, Department of  Physics, 119991
Moscow, Russia\\~\\ P.N. Lebedev Physical institute of RAS, 119991,
Moscow, Russia}

\date{\today }
7 pages, 3 figures
\begin{abstract}
We investigated the peculiarities of non-equilibrium charge
configurations in the system of two strongly coupled quantum dots
(QDs) weakly connected to the reservoirs in the presence of Coulomb
correlations. We revealed that total electron occupation
demonstrates in some cases significant decreasing with increasing of
applied bias - contrary to the situation when Coulomb correlations
are absent and found well pronounced ranges of system parameters
where negative tunneling conductivity appears due to the Coulomb
correlations.
\end{abstract}

\maketitle

\section{Introduction}

Electron tunneling through the system of coupled QDs in the presence
of strong Coulomb correlations is one of the most interesting
problems in the solid state physics. The present day experimental
technique gives possibility to produce single QDs with a given set
of parameters and to create coupled QDs with different spatial
geometry \cite{Vamivakas},\cite{Stinaff},\cite{Munoz-Matutano}.
Thereby the main effort in the physics of QDs is devoted to the
investigation of non-equilibrium charge states and different charge
configurations due to the electrons tunneling
\cite{Kikoin_1},\cite{Goldin},\cite{Arseyev_1},\cite{Arseyev_2},\cite{Arseyev_3}
through the system of coupled QDs in the presence of strong Coulomb
interaction.

One of the most intensively studied problems in this field is
tunneling through the single QD \cite{Paaske_1},\cite{Kaminski} and
interacting QDs \cite{Kikoin_1},\cite{Goldin},\cite{Orellana},
\cite{Lopez} in the Kondo regime, which reveals rich physics for
small bias voltage compared to the tunneling rates. It was
demonstrated \cite{Orellana} that Coulomb correlations in QDs lead
to bistable behavior in the Kondo regime at zero bias voltage.
Charge redistribution between different spin configurations in the
system of two interacting QDs in the Kondo regime was regarded in
\cite{Kikoin_1}. The authors considered the situation when the
detuning between the energy levels in the QDs exceeds the dots
coupling and on-site Coulomb repulsion is present only in a single
dot.

A great attention is also payed to the double QDs as attractive
systems for spin-dependent transport
\cite{Golovach},\cite{Hornberger},\cite{Fransson}. In
\cite{Golovach} authors studied transport through double QD both in
sequential tunneling and co-tunneling regimes by means of master
equation for density matrix in the basis of exact eigenfunctions and
eigenvalues. Unfortunately only transitions between the empty states
and states with one and two electrons were considered. Results
obtained in \cite{Hornberger} and \cite{Fransson} deal with the
investigation of transport in the a double QD system weakly coupled
to spin-polarized leads. The method of charge and spin transport
analysis presented in \cite{Hornberger} is based on the Liouville
equation for the reduced density matrix in lowest order in the
tunneling transitions. Authors analyzed tunneling conductivity and
$I-V$ characteristics as a functions of magnetic leads polarization
and gate voltage. The system where only one or two electrons can be
localized simultaneously due to the specific features of the
ferromagnetic leads was investigated in \cite{Fransson}.

In the present paper we consider electron tunneling through the
coupled QDs in the regime when applied bias can be tuned in a wide
range and the on-site Coulomb repulsion can be comparable to the
other system parameters. We analyze different charge configurations
in the system of two strongly coupled quantum dots (QDs) weakly
connected to the reservoirs in the presence of Coulomb correlations
in a wide range of applied bias in terms of pseudo operators with
constraint
\cite{Coleman},\cite{Coleman_1},\cite{Wingreen},\cite{Arseyev_3}.
For large values of applied bias Kondo effect is not essential so we
neglect any correlations between electron states in the QDs and in
the leads. This approximation allows to describe correctly non
equilibrium occupation of any single- and multi-electron state due
to  the tunneling processes. We revealed the presence of negative
tunneling conductivity in certain ranges of the applied bias voltage
and revealed that total electron occupation demonstrates in some
cases significant decreasing with increasing of applied bias -
contrary to the situation when Coulomb correlations are absent.

\section{Model}
We consider a system of coupled QDs with the single particle levels
$\widetilde{\varepsilon}_{1}$ è $\widetilde{\varepsilon}_{2}$
connected to the two leads. The Hamiltonian can be written as:

\begin{eqnarray}
\hat{H}&=&\sum_{\sigma}c_{1\sigma}^{+}c_{1\sigma}\widetilde{\varepsilon}_{1}+
\sum_{\sigma}c_{2\sigma}^{+}c_{2\sigma}\widetilde{\varepsilon}_{2}+
U_1\widehat{n}_{1\sigma}\widehat{n}_{1-\sigma}+\nonumber\\&+&U_2\widehat{n}_{2\sigma}\widehat{n}_{2-\sigma}+
\sum_{\sigma}T(c_{1\sigma}^{+}c_{2\sigma}+c_{2\sigma}^{+}c_{1\sigma})
\end{eqnarray}

where operator $c_{l\sigma}$ creates an electron in the dot $i$ with
spin $\sigma$, $\widetilde{\varepsilon}_{l}$ is the energy of the
single electron level in the dot $i$ and $T$ is the inter-dot
tunneling coupling, $n_{l\sigma}=c_{l\sigma}^{+}c_{l\sigma}$ and
$U_{1(2)}$ is the on-site Coulomb repulsion of localized electrons.
When the coupling between QDs exceeds the value of interaction with
the leads, one has to use the basis of exact eigenfunctions and
eigenvalues of coupled QDs without the interaction with the leads.
In this case all energies of single- and multi-electron states are
well known:

One electron in the system: two single electron states with the wave
function

\begin{eqnarray}
\psi_{i}^{\sigma}=\mu_{i}\cdot|0\uparrow\rangle|00\rangle+\nu_{i}\cdot|00\rangle|0\uparrow\rangle
\end{eqnarray}
Single electron energies and coefficients $\mu_{i}$ and $\nu_{i}$
can be found as an eigenvalues and eigenvectors of matrix:

\begin{eqnarray}
\begin{pmatrix}\varepsilon_{1} && -T\\
-T && \varepsilon_{2}\end{pmatrix} \label{m1}\end{eqnarray}

Two electrons in the system: two states with the same spin
$\sigma\sigma$ and $-\sigma-\sigma$ and four two-electron states
with the opposite spins $\sigma-\sigma$ with the wave function:

\begin{eqnarray}
\psi_{j}^{\sigma-\sigma}&=&\alpha_{j}\cdot|\uparrow\downarrow\rangle|00\rangle+\beta_{k}\cdot|\downarrow0\rangle|0\uparrow\rangle+\nonumber\\&+&
\gamma_{j}\cdot|0\uparrow\rangle|\downarrow0\rangle+\delta_{j}\cdot|00\rangle|\uparrow\downarrow\rangle\nonumber\\
\end{eqnarray}

Two electron energies and coefficients $\alpha_{j}$, $\beta_{j}$,
$\gamma_{j}$ and $\delta_{j}$ are the eigenvalues and eigenvectors
of matrix:

\begin{eqnarray}
\begin{pmatrix}2\varepsilon_{1}+U_{1} && -T && -T && 0 \\
-T && \varepsilon_{1}+\varepsilon_{2} && 0 && -T\\
-T && 0 && \varepsilon_{1}+\varepsilon_{2} && 0\\
0 && -T && -T && 2\varepsilon_{2}+U_{2}\end{pmatrix}
\label{m2}\end{eqnarray}

Three electrons in the system: two three-electron states with the
wave function

\begin{eqnarray}
\psi_{m}^{\sigma\sigma-\sigma}&=&p_{m}|\uparrow\downarrow\rangle|\uparrow\rangle+q_{m}|\uparrow\rangle|\uparrow\downarrow\rangle\nonumber\\
m&=&\pm1
\end{eqnarray}

Three electron energies and coefficients $p_{m}$ and $Q_{m}$ can be
found as an eigenvalues and eigenvectors of matrix:

\begin{eqnarray}
\begin{pmatrix}2\varepsilon_{1}+\varepsilon_{2}+U_{1} && -T\\
-T && 2\varepsilon_{2}+\varepsilon_{1}+U_{2}\end{pmatrix}
\label{m3}\end{eqnarray}

Four electrons in the system: one four-electron state with energy
$E_{IVl}=2\varepsilon_1+2\varepsilon_2+U_1+U_{2}$ and wave function

\begin{eqnarray}
\psi_{l}=|\uparrow\downarrow\rangle|\uparrow\downarrow\rangle
\end{eqnarray}

If coupled QDs are connected with the leads of the tunneling contact
the number of electrons in the dots changes due to the tunneling
processes. Transitions between the states with different number of
electrons in the two interacting QDs can be analyzed in terms of
pseudo-particle operators with constraint on the physical states
(the number of pseudo-particles). Consequently, the electron
operator $c_{\sigma l}^{+}$  $(l=1,2)$ can be written in terms of
pseudo-particle operators as:

\begin{eqnarray}
c_{\sigma l}^{+}&=&\sum_{i}X_{i}^{\sigma l}f_{\sigma
i}^{+}b+\sum_{j,i}Y_{ji}^{\sigma-\sigma
l}d_{j}^{+\sigma-\sigma}f_{i-\sigma}+\\&+&\sum_{j,i}Y_{i}^{\sigma\sigma
l}d^{+\sigma\sigma}f_{i\sigma}+\sum_{m,j}Z_{mj}^{\sigma\sigma-\sigma
l}\psi_{m-\sigma}^{+}d_{j}^{\sigma-\sigma}+\nonumber\\&+&\sum_{m}Z_{m}^{\sigma-\sigma-\sigma
l}\psi_{m\sigma}^{+}d^{-\sigma-\sigma}+\sum_{m}W_{m}^{\sigma-\sigma-\sigma
l}\varphi^{+}\psi_{m\sigma}\nonumber\
\end{eqnarray}

where $f_{\sigma}^{+}(f_{\sigma})$ and
$\psi_{\sigma}^{+}(\psi_{\sigma})$- are pseudo-fermion creation
(annihilation) operators for the electronic states with one and
three electrons correspondingly. $b^{+}(b)$,
$d_{\sigma}^{+}(d_{\sigma})$ and $\varphi^{+}(\varphi)$- are slave
boson operators, which correspond to the states without any
electrons, with two electrons or four electrons. Operators
$\psi_{m-\sigma}^{+}$- describe system configuration with two spin
up electrons $\sigma$ and one spin down electron $-\sigma$ in the
symmetric and asymmetric states.

Matrix elements $X_{i}^{\sigma l}$, $Y_{ji}^{\sigma-\sigma l}$,
$Y_{ji}^{\sigma\sigma l}$, $Z_{mj}^{\sigma\sigma-\sigma l}$,
$Z_{mj}^{\sigma-\sigma-\sigma l}$ and $W_{m}^{\sigma-\sigma-\sigma
l}$ can be evaluated as:

\begin{eqnarray}
X_{i}^{\sigma l}&=&\langle\psi_{i}^{\sigma}|c_{\sigma l}^{+}|0\rangle\nonumber\\
Y_{ji}^{\sigma-\sigma l}&=&\langle\psi_{j}^{\sigma-\sigma}|c_{\sigma il}^{+}|\psi_{i}^{-\sigma}\rangle\nonumber\\
Y_{ji}^{\sigma\sigma l}&=&\langle\psi_{j}^{\sigma\sigma}|c_{\sigma l}^{+}|\psi_{i}^{\sigma}\rangle\nonumber\\
Z_{mj}^{\sigma\sigma-\sigma l}&=&\langle\psi_{m}^{\sigma\sigma-\sigma}|c_{\sigma l}^{+}|\psi_{j}^{\sigma-\sigma}\rangle\nonumber\\
Z_{m}^{\sigma-\sigma-\sigma l}&=&\langle\psi_{m}^{\sigma-\sigma-\sigma}|c_{\sigma l}^{+}|\psi^{-\sigma-\sigma}\rangle\nonumber\\
W_{m}^{\sigma-\sigma-\sigma
l}&=&\langle\psi_{l}^{\sigma\sigma-\sigma-\sigma}|c_{\sigma
l}^{+}|\psi_{m}^{\sigma-\sigma-\sigma}\rangle\
\end{eqnarray}

Finally one can easily express matrix elements through the matrixes
(\ref{m1}), (\ref{m2}), (\ref{m3}) eigenvectors elements:

\begin{eqnarray}
X_{i}^{\sigma 1}=\mu_{i};
X_{i}^{\sigma 2}=\nu_{i}\nonumber\\
Y_{ji}^{\sigma-\sigma 1}=\alpha_j\mu_i+\beta_j\nu_i\nonumber\\
Y_{ji}^{\sigma-\sigma 2}=\delta_j\nu_i+\gamma_j\mu_i\nonumber\\
Y_{ji}^{\sigma\sigma 1}=\nu_i;
Y_{ji}^{\sigma\sigma 2}=\mu_i\nonumber\\
Z_{mj}^{\sigma\sigma-\sigma 1}=p_m\gamma_j+q_m\delta_j\nonumber\\
Z_{mj}^{\sigma\sigma-\sigma 2}=p_m\alpha_j+q_m\beta_j\nonumber\\
Z_{mj}^{\sigma-\sigma-\sigma 1}=p_m;
Z_{mj}^{\sigma-\sigma-\sigma 1}=q_m\nonumber\\
W_{m}^{\sigma-\sigma-\sigma 1}=q_m; W_{m}^{\sigma-\sigma-\sigma
2}=p_m
\end{eqnarray}

The constraint on the space of the possible system states have to be
taken into account:

\begin{eqnarray}
\widehat{n}_{b}+\sum_{i\sigma}\widehat{n}_{fi\sigma}+\sum_{j\sigma\sigma^{'}}\widehat{n}_{dj}^{\sigma\sigma^{'}}+\sum_{m\sigma}\widehat{n}_{\psi
m\sigma}+\widehat{n}_{\varphi}=1 \label{limit}
\end{eqnarray}

Condition (\ref{limit}) means that the appearance  of any two
pseudo-particles in the system simultaneously is impossible.

Electron filling numbers in the coupled QDs can be expressed in the
terms of the pseudo-particles filling numbers:

\begin{eqnarray}
\widehat{n}_{\sigma}^{el}&=&\sum_{l}c_{\sigma l}^{+}c_{\sigma
l}=\sum_{i,l}|X_{i}^{\sigma
l}|^{2}\widehat{n}_{fi\sigma}+\sum_{i,j,l}|Y_{ji}^{\sigma-\sigma
l}|^{2}\widehat{n}_{dj}^{\sigma-\sigma}+\nonumber\\&+&\sum_{i,l}|Y_{ji}^{\sigma\sigma
l}|^{2}\widehat{n}_{dj}^{\sigma\sigma}+\sum_{m,j,l}|Z_{mj}^{\sigma\sigma-\sigma
l}|^{2}\widehat{n}_{\psi
m-\sigma}+\nonumber\\&+&\sum_{m,l}|Z_{mj}^{-\sigma-\sigma\sigma
l}|^{2}\widehat{n}_{\psi
m\sigma}+\sum_{m,l}|W_{m}^{\sigma-\sigma-\sigma
l}|^{2}\widehat{n}_{\varphi}
\end{eqnarray}

Consequently, the Hamiltonian of the system can be written in the
terms of the pseudo-particle operators:

\begin{eqnarray}
\hat{H}&=&\hat{H_{0}}+\hat{H}_{tun}\\
\hat{H_{0}}&=&\sum_{i\sigma}\varepsilon_{i}f_{i\sigma}^{+}f_{i\sigma}+\sum_{j\sigma\sigma^{'}}E_{IIj}^{\sigma\sigma^{'}}d_{j}^{+\sigma\sigma^{'}}d_{j}^{\sigma\sigma^{'}}+\nonumber\\&+&\sum_{m\sigma}E_{III}^{m\sigma}\psi_{m\sigma}^{+}\psi_{m\sigma}+E_{IVl}\varphi_{\sigma}^{+}\varphi_{\sigma}+\nonumber\\&+&\sum_{k\sigma}(\varepsilon_{k\sigma}-eV)c_{k\sigma}^{+}c_{k\sigma}+\sum_{p\sigma}\varepsilon_{p\sigma}c_{p\sigma}^{+}c_{p\sigma}\nonumber\\
\hat{H}_{tun}&=&\sum_{k\sigma}T_{k}(c_{k\sigma}^{+}c_{\sigma1}+c_{\sigma1}^{+}c_{k\sigma})+(k\leftrightarrow
p;1\leftrightarrow2)\nonumber\
\end{eqnarray}

where $\varepsilon_i$, $E_{IIj}^{\sigma\sigma^{'}}$,
$E_{III}^{m\sigma}$ and $E_{IVl}$-are the energies of the single-,
double-, triple- and quadri-electron states.
$\varepsilon_{k(p)\sigma}$-is the energy of the conduction electrons
in the states $k$ and $p$ correspondingly.
$c_{k(p)\sigma}^{+}/c_{k(p)\sigma}$ are the creation (annihilation)
operators in the leads of the tunneling contact. $T_{k(p)}$-are the
tunneling amplitudes, which we assume to be independent on momentum
and spin. Indexes $k(p)$ mean only that tunneling takes place from
the system of coupled QDs to the conduction electrons in the states
$k$ and $p$ correspondingly.

Bilinear combinations of pseudo-particle operators are closely
connected with the density matrix elements. So, similar expressions
can be obtained from equations for the density matrix evolution but
method based on the pseudo particle operators is more compact and
convenient. The tunneling current through the proposed system
written in terms of the pseudo-particle operators has the form:

\begin{eqnarray}
\widehat{I}_{k\sigma}&=&\sum_{k}\frac{\partial
\widehat{n}_{k}}{\partial t}=i [\sum_{ik}X_{i}^{\sigma 1}T_{k}
c_{k\sigma}f_{i\sigma}^{+}b+\nonumber\\&+&\sum_{ijk}Y_{ji}^{\sigma-\sigma
1}T_{k}
c_{k\sigma}d_{j}^{+\sigma-\sigma}f_{i-\sigma}+\sum_{ijk}Y_{ji}^{\sigma\sigma
1}T_{k} c_{k\sigma}d_{j}^{+\sigma\sigma}f_{i\sigma}+\nonumber\\&+&
\sum_{mjk}Z_{mj}^{\sigma\sigma-\sigma
1}T_{k}c_{k\sigma}\psi_{m-\sigma}^{+}d_{j}^{\sigma-\sigma}+\nonumber\\
&+& \sum_{mjk}Z_{mj}^{-\sigma-\sigma\sigma
1}T_{k}c_{k\sigma}\psi_{m\sigma}^{+}d_{j}^{-\sigma-\sigma}+\nonumber\\&+&\sum_{mk}W_{m}^{\sigma-\sigma-\sigma
1}T_{k}c_{k\sigma}\varphi^{+}\psi_{m\sigma}-h.c.]
\end{eqnarray}

We set $\hbar=1$ and neglect changes in the electron spectrum and
local density of states in the tunneling contact leads, caused by
the tunneling current. Therefore equations of motion together with
the constraint on the space of the possible system states
(pseudo-particles number) (\ref{limit}) give the following
equations:

\begin{eqnarray}
Im\sum_{ik}T_{k}X_{i}^{\sigma 1}\cdot \langle
c_{k\sigma}f_{i\sigma}^{+}b\rangle=\nonumber\\
=\Gamma_{k}\sum_{i}[(1-n_{k\sigma}(\varepsilon_i))\cdot
n_{fi\sigma}-n_{k\sigma}(\varepsilon_{i})\cdot n_{b}](X_{i}^{\sigma 1})^{2}\nonumber\\
Im\sum_{ijk}Y_{ji}^{\sigma-\sigma 1}T_{k}\cdot \langle
c_{k\sigma}d_{j}^{+\sigma-\sigma}f_{i-\sigma}\rangle=\nonumber\\=
\Gamma_{k}\sum_{ij}[(1-n_{k\sigma}(E_{IIj}^{\sigma-\sigma}-\varepsilon_{i-\sigma}))\cdot
n_{dj}^{\sigma-\sigma}-\nonumber\\-n_{k\sigma}(E_{IIj}^{\sigma-\sigma}-\varepsilon_{i-\sigma})\cdot
n_{fi-\sigma}](Y_{ji}^{\sigma-\sigma
1})^{2}\nonumber\\
Im\sum_{ijk}Y_{ji}^{\sigma\sigma 1}T_{k}\cdot \langle
c_{k\sigma}d_{j}^{+\sigma\sigma}f_{i\sigma}\rangle=\nonumber\\=
\Gamma_{k}\sum_{ij}[(1-n_{k\sigma}(E_{IIj}^{\sigma\sigma}-\varepsilon_{i\sigma}))\cdot
n_{dj}^{\sigma\sigma}-\nonumber\\-n_{k\sigma}(E_{IIj}^{\sigma\sigma}-\varepsilon_{i\sigma})\cdot
n_{fi\sigma}](Y_{ji}^{\sigma\sigma
1})^{2}\nonumber\\
Im\sum_{mjk}Z_{mj}^{\sigma\sigma-\sigma 1} T_{k}\cdot \langle
c_{k\sigma}\psi_{m-\sigma}^{+}d_{j}^{\sigma-\sigma}\rangle=\nonumber\\=
\Gamma_{k}\sum_{mj}[(1-n_{k\sigma}(E_{III}^{m-\sigma}-E_{IIj}^{\sigma-\sigma}))\cdot
n_{\psi
m-\sigma}-\nonumber\\-n_{k\sigma}(E_{III}^{m-\sigma}-E_{IIj}^{\sigma-\sigma})\cdot
n_{dj}^{\sigma-\sigma}](Z_{mj}^{\sigma\sigma-\sigma
1})^{2}\nonumber\\
Im\sum_{mjk}Z_{mj}^{-\sigma-\sigma\sigma 1} T_{k}\cdot \langle
c_{k\sigma}\psi_{m\sigma}^{+}d_{j}^{-\sigma-\sigma}\rangle=\nonumber\\=
\Gamma_{k}\sum_{mj}[(1-n_{k\sigma}(E_{III}^{m\sigma}-E_{IIj}^{-\sigma-\sigma}))\cdot
n_{\psi
m\sigma}-\nonumber\\-n_{k\sigma}(E_{III}^{m\sigma}-E_{IIj}^{-\sigma-\sigma})\cdot
n_{dj}^{-\sigma-\sigma}](Z_{mj}^{-\sigma-\sigma\sigma
1})^{2}\nonumber\\
Im\sum_{mk}W_{m}^{\sigma-\sigma-\sigma 1} T_{k}\cdot \langle
c_{k\sigma}\varphi_{l}^{+}\psi_{m\sigma}\rangle=\nonumber\\
\Gamma_{k}\sum_{m}[(1-n_{k\sigma}(E_{IVl}-E_{III}^{m\sigma}))\cdot
n_{\varphi}-\nonumber\\-n_{k\sigma}(E_{IVl}-E_{III}^{m\sigma})\cdot
n_{\psi m\sigma}](W_{m}^{\sigma-\sigma-\sigma 1})^{2}
\label{tunneling_current}
\end{eqnarray}

Tunneling current $I_{k\sigma}$ is determined by the sum of the
right hand parts of the equations (\ref{tunneling_current}).

Stationary system of equations can be obtained for the pseudo
particle filling numbers  $n_{fi}$, $n_{dj}^{\sigma-\sigma}$,
$n_{d}^{\sigma\sigma}$, $n_{\psi m}$ and $n_{\varphi}$:

\begin{eqnarray}
0&=&\frac{\partial n_{\varphi}}{\partial t}=-\Gamma_{k}\sum_{m\sigma
}[-n_{\psi m\sigma}\cdot
n_{k\sigma}(E_{IVl}-E_{III}^{m\sigma})+\nonumber\\&+&n_{\varphi}\cdot(1-n_{k\sigma}(E_{IVl}-E_{III}^{m\sigma}))]|W_{m}^{\sigma-\sigma-\sigma 1}|^{2}+(k,1\leftrightarrow p,2)\nonumber\\
0&=&\frac{\partial n_{\psi m\sigma}}{\partial
t}=-\Gamma_{k}\sum_{j}[n_{\psi m\sigma}\cdot
(1-n_{k-\sigma}(E_{III}^{m\sigma}-E_{IIj}^{\sigma-\sigma}))-\nonumber\\&-&n_{k-\sigma}(E_{III}^{m\sigma}-E_{IIj}^{\sigma-\sigma})\cdot
n_{dj}^{\sigma-\sigma}]|(Z_{mj}^{\sigma\sigma-\sigma
1}|^{2}-\nonumber\\
&-&\Gamma_{k}\sum_{j}[(1-n_{k\sigma}(E_{III}^{m\sigma}-E_{IIj}^{-\sigma-\sigma}))\cdot
n_{\psi m\sigma}-\nonumber\\&-&n_{dj}^{-\sigma-\sigma}\cdot
n_{k\sigma}(E_{III}^{m\sigma}-E_{IIj}^{-\sigma-\sigma})]|Z_{mj}^{\sigma-\sigma-\sigma
1}|^{2}-\nonumber\\
&-&\Gamma_{k}[-(1-n_{k\sigma}(E_{IVl}-E_{III}^{m\sigma}))\cdot
n_{\varphi}+\nonumber\\&+&n_{\psi m\sigma}\cdot
n_{k\sigma}(E_{IVl}-E_{III}^{m\sigma})]|W_{m}^{\sigma-\sigma-\sigma
1}|^{2}+(k,1\leftrightarrow p,2)\nonumber\\
0&=&\frac{\partial n_{dj}^{\sigma\sigma}}{\partial
t}=-\Gamma_{k}\sum_{i}[(1-n_{k\sigma}(E_{IIj}^{\sigma\sigma}-\varepsilon_{i}))\cdot
n_{dj}^{\sigma\sigma}-\nonumber\\&-&n_{k\sigma}(E_{IIj}^{\sigma\sigma}-\varepsilon_{i})\cdot
n_{fi\sigma}]|Y_{ji}^{\sigma\sigma
1}|^{2}-\nonumber\\
&-&\Gamma_{k}\sum_{m}[n_{k-\sigma}(E_{III}^{m-\sigma}-E_{IIj}^{\sigma\sigma})\cdot
n_{dj}^{\sigma\sigma}-\nonumber\\&-&(1-n_{k-\sigma}(E_{III}^{m-\sigma}-E_{IIj}^{\sigma\sigma}))\cdot
n_{\psi m-\sigma}]|Z_{mj}^{\sigma-\sigma-\sigma 1}|^{2}+\nonumber\\&+&(k,1\leftrightarrow p,2)\nonumber\\
0&=&\frac{\partial n_{dj}^{\sigma-\sigma}}{\partial
t}=-\Gamma_{k}\sum_{i\sigma}[(1-n_{k-\sigma}(E_{IIj}^{\sigma-\sigma}-\varepsilon_{i}))\cdot
n_{dj}^{\sigma-\sigma}-\nonumber\\&-&n_{k-\sigma}(E_{IIj}^{\sigma-\sigma}-\varepsilon_{i})\cdot
n_{fi\sigma}]|Y_{ji}^{\sigma-\sigma
1}|^{2}-\nonumber\\&-&\Gamma_{k}\sum_{m\sigma}[n_{k\sigma}(E_{III}^{m\sigma}-E_{IIj}^{\sigma-\sigma})\cdot
n_{dj}^{\sigma-\sigma}-\nonumber\\&-&(1-n_{k\sigma}(E_{III}^{m\sigma}-E_{IIj}^{\sigma-\sigma}))\cdot
n_{\psi m-\sigma}]|Z_{mj}^{\sigma\sigma-\sigma
1}|^{2}+\nonumber\\&+&(k,1\leftrightarrow p,2)\nonumber\\
0&=&\frac{\partial n_{fi\sigma}}{\partial
t}=\Gamma_{k}[n_{k\sigma}(\varepsilon_i)\cdot
n_{b}-(1-n_{k\sigma}(\varepsilon_i))\cdot
n_{fi\sigma}]|X_{i}^{\sigma1}|^{2}+\nonumber\\
&+&\Gamma_{k}\sum_{j\sigma}[(1-n_{k-\sigma}(E_{IIj}^{\sigma-\sigma}-\varepsilon_i))\cdot
n_{dj}^{\sigma-\sigma}-\nonumber\\&-&n_{k-\sigma}(E_{IIj}^{\sigma-\sigma}-\varepsilon_i)\cdot
n_{fi\sigma}]|Y_{ji}^{\sigma-\sigma
1}|^{2}+\nonumber\\
&+&\Gamma_{k}\sum_{j}[(1-n_{k\sigma}(E_{IIj}^{\sigma\sigma}-\varepsilon_{i}))\cdot
n_{dj}^{\sigma\sigma}-\nonumber\\&-&n_{k\sigma}(E_{IIj}^{\sigma\sigma}-\varepsilon_{i})\cdot
n_{fi\sigma}]|Y_{ji}^{\sigma\sigma 1}|^{2}+(k,1\leftrightarrow
p,2)\nonumber\ 
\end{eqnarray}

In these equations we neglect the non-diagonal averages of
pseudo-particle operators such as $\langle
f_{\sigma}^{+}bf_{-\sigma}^{+}d\rangle$ etc.. These terms are of the
next order in small parameter $\Gamma_{k(p)}/\Delta E$ where $\Delta
E$ is the energy difference between any energy states in the coupled
QDs. We consider the paramagnetic situation, when conditions
$n_{fi\sigma}=n_{fi-\sigma}$, $n_{\psi m\sigma}=n_{\psi m-\sigma}$,
$n_{k\sigma}=n_{k-\sigma}$ and
$n_{dj}^{-\sigma-\sigma}=n_{dj}^{\sigma\sigma}$ are fulfilled.
System of equations (\ref{sys1}) in the stationary case is the
linear system, which allows to determine pseudo particle filling
numbers, electron filling numbers $n_{el}(eV)$ and tunneling current
$I_{k\sigma}$.

\section{Results and discussion}

\begin{figure} [h!]
\includegraphics[width=65mm]{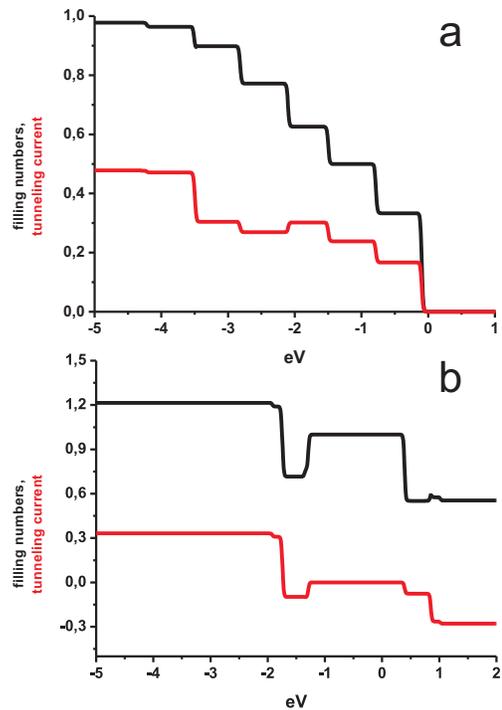}
\caption{Fig.1 (Color online) Coupled QDs filling numbers (black
line) and tunneling current (red line) as a functions of applied
bias voltage in the case of symmetrical tunneling contact
$\Gamma_{k}=\Gamma_{p}=0.01$. a).
$\varepsilon_{1}=\varepsilon_{2}=0.8$, $T=0.7$, $U_{1}=U_{2}=2.0$;
b). $\varepsilon_{1}=-0.5$, $\varepsilon_{2}=-0.6$, $T=0.3$,
$U_{1}=U_{2}=2.0$ } \label{Fig.1}
\end{figure}

\begin{figure*} [h!]
\includegraphics[width=160mm]{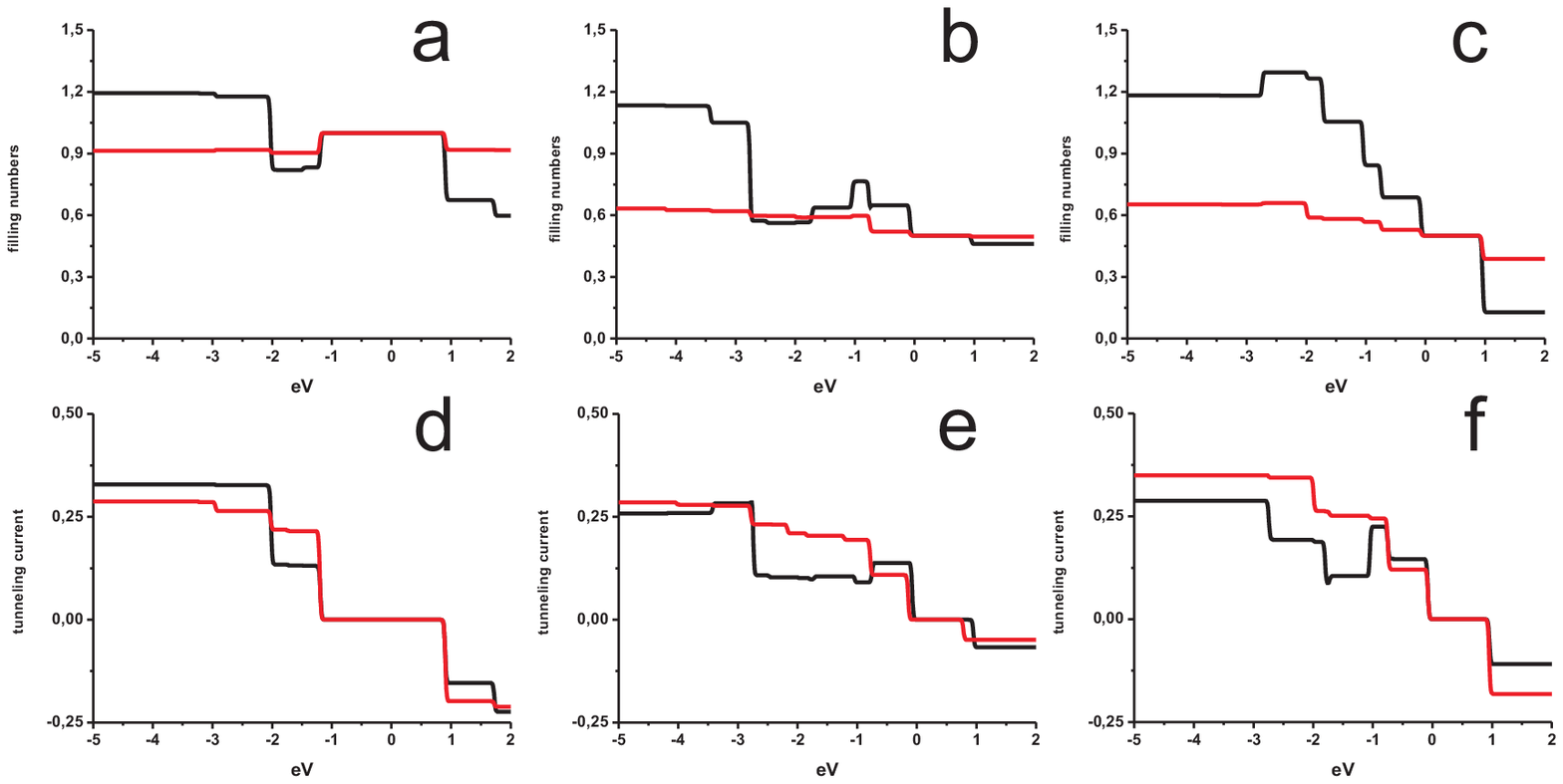}
\caption{Fig.2 (Color online) Coupled QDs filling numbers a).-c).
and tunneling current d).-f). as a functions of applied bias voltage
in the case of symmetrical $\Gamma_{k}=\Gamma_{p}=0.01$ (black line)
and asymmetrical $\Gamma_{k}=0.01$, $\Gamma_{p}=0.1$ (red line)
tunneling contact. Parameters$U_{1}=2.0$ and $U_{2}=2.0$ are the
same for all the figures. a),d).$\varepsilon_1=-0.5$,
$\varepsilon_{2}=-1.2$, $T=0.8$; b),e).$\varepsilon_1=-0.5$,
$\varepsilon_{2}=-0.7$, $T=0.6$; c),f).$\varepsilon_1=-0.7$,
$\varepsilon_{2}=-0.5$, $T=0.6$.} \label{Fig.2}
\end{figure*}

The behavior of the total electron occupation of the coupled QDs
$n_{el}(eV)$ and $I-V$ characteristics are presented on the
Fig.\ref{Fig.1} and Fig.\ref{Fig.2}. We first analyze the behavior
of the the total electron occupation of the QDs $n_{el}(eV)$ and
$I-V$ characteristics of the considered system for different single
electron levels positions relative to the sample Fermi level and
various tunneling rates to the contact leads. The bias voltage in
our calculations is applied to the sample. Consequently, if both
single electron levels are above(below) the Fermi level, all the
specific features of the total electron occupation and tunneling
current characteristics can be observed at negative(positive) values
of $eV$. In the case when both single electron energy levels are
situated above the sample Fermi level [Fig.\ref{Fig.1}a (black
line)] we observe the step-like behavior of the total electron
occupation. The width and height of the steps are determined by the
relation between the system parameters $T$, $\varepsilon$ and $U$
and $\Gamma=\Gamma_k+\Gamma_p$. The tunneling current is depicted in
[Fig.\ref{Fig.1}a (red line)] as a function of the applied bias. It
is clearly evident that even for symmetric tunneling contact the
presence of Coulomb interaction leads to the appearance of negative
tunneling conductivity. Negative tunneling conductivity was found in
the double QDs system connected to the magnetic leads only for
particular polarization direction \cite{Hornberger},\cite{Fransson}.
We also want to point out that negative tunneling conductivity in
coupled QDs can observed experimentally \cite{Liu}.

The most interesting result were obtained in the case when both
single electron energy levels in the QDs are situated below the
Fermi level (Fig.\ref{Fig.1}b (black line), Fig.\ref{Fig.2}a-c). In
the particular range of the applied bias both for the symmetric
$\Gamma_k=\Gamma_p$ and asymmetric $\Gamma_k>\Gamma_p$ tunneling
contact the total electron occupation demonstrate significant jumps.
In a QD without Coulomb interaction the total system occupation can
only increase as applied bias increases and passes single electron
levels. Quite different situation occurs in a system with strong
Coulomb correlations. For zero bias electron filling numbers are
determined by equilibrium occupation. But when the increasing bias
reaches the energies of multi-electron excited states, the total
occupation begins to decrease. Using single-electron language we can
say that additional tunneling electrons "push out" electrons from
the states below the Fermi level due to Coulomb repulsion. Or one
can look at this effect as increasing of probability for electrons
to leave the QD due to appearance of several non-elastic channels of
tunneling (accompanied with changing of multi-electron states of the
QD).

The tunneling current as a function of the applied bias for this
case is depicted in Figures \ref{Fig.1}b (red line),\ref{Fig.2}d-f
and reveals not only the monotonic step-like behavior, but also
demonstrates the appearance of negative tunneling conductivity.

\begin{figure} [h!]
\includegraphics[width=65mm]{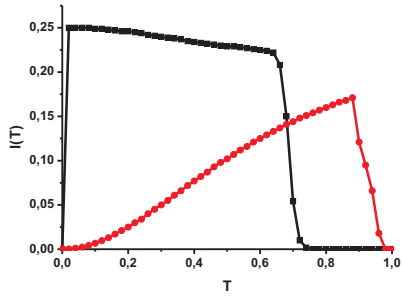}
\caption{Fig.3 (Color online) Tunneling current as a functions of
Rabi frequency in the case of symmetrical tunneling contact in the
presence (black line) and in the absence (red line) of Coulomb
interaction. $\varepsilon_{1}=-0.7$, $\varepsilon_{2}=0.5$,
$eV=-1.0$ $U_{1}=U_{2}=2.0$,$\Gamma_{k}=\Gamma_{p}=0.01$ }
\label{Fig.3}
\end{figure}

Our model also describes the situation, when two QDs (or two levels
in one QD) are coupled due to interaction with external field in
rotating wave approximation and direct tunneling coupling is
negligible. In this case Hamiltonian, which describes interaction
with external field, has the form:

\begin{eqnarray}
\hat{H}_{int}&=&=\sum_{\sigma,\beta}(d_{\beta})_{12}\epsilon_{\beta}c_{1\sigma}^{+}c_{2\sigma}+h.c.
\end{eqnarray}

where $(d_{\beta})_{12}$ - are matrix elements for dipole
transitions, $\epsilon_{\beta}$ - external field components and
$\beta=x,y,z$. So, all the calculations and results remain valid if
tunneling transfer amplitude $T$ is replaced by
$\sum_{\beta}(d_{\beta})_{12}\epsilon_{\beta}=\frac{\Omega}{2}$,
where $\Omega$ is a Rabi frequency. Consequently the dependence of
tunneling current on the external field intensity at the fixed value
of applied bias can be obtained from $I(\Omega)$ (Fig.\ref{Fig.3}).
In the presence of Coulomb interaction the dependence of tunneling
current on external field intensity reveals fast switching on and
off with the external field intensity increasing. Between the
switchings tunneling current amplitude remains nearly constant (see
black line in Fig.\ref{Fig.3}). In the absence of Coulomb
interaction the tunneling current amplitude smoothly increases with
the external field intensity growth. Tunneling current is rapidly
switched off at the critical intensity value, which exceeds the
critical switching off value for non-zero Coulomb interaction. So,
the proposed system can be applied as an effective switch key.

\section{Conclusion}

We investigated tunneling through the system of two strongly coupled
QDs weakly interacting with the reservoirs in the presence of
Coulomb correlations between localized electrons. In the considered
system if electrons number changes due to the tunneling processes,
the modification of the energy spectrum is not reduced to the simple
adding of Coulomb interaction $U$ per electron. One, two, three or
four electrons can be localized in the coupled QDs , each state with
fixed total charge and spin projection has it's own energy.
Transitions between these states were analyzed in terms of
pseudo-particle operators with constraint on the possible physical
states of the system. Filling numbers of different multi-electron
states, total electron occupation of QDs and $I-V$ characteristics
were investigated for different single electron levels positions
relative to the sample Fermi level and various tunneling transfer
rates.

It was shown that total electron occupation can significantly
decrease with increasing of applied bias when both single electron
energy levels are situated below the Fermi level contrary to the
situation with no correlations. Tunneling electron changes
multi-particle states of QDs and pushes out electrons from the state
below the Fermi level. We revealed that for some parameter range,
the system demonstrates negative tunneling conductivity in certain
ranges of the applied bias voltage due to the Coulomb correlations.

We demonstrated, that correlated QDs can be used as an effective
current switch key by changing the intensity of the external field.

This work was partly supported by the RFBR and  Leading Scientific
School grants. The support from the Ministry of Science and
Education is also acknowledged.


\pagebreak

\end{document}